\documentclass[11pt,twoside]{article}

\usepackage{asp2006}

\usepackage{graphicx}

\begin{document}
\title{Transient Mass-Loss Events in the PG~1159 Central Star of Longmore~4}   
\author{Howard E. Bond}   
\affil{Space Telescope Science Institute, Baltimore, MD USA}    

\begin{abstract} 
A spectacular, transient mass-loss event in the hydrogen-deficient central star
of the planetary nebula Longmore~4 was discovered in 1992 by Werner et al. 
During that event, the star temporarily changed from its normal PG~1159 spectrum
to that of an emission-line [WCE] star. I have been monitoring the spectrum of
Lo~4 since 2003.  Out of 31 spectra, two of them reveal mass-loss outbursts
similar to the one seen in 1992, showing that the phenomenon recurs. I speculate
on possible mechanisms for these unique  outbursts, but emphasize that we still
have no fully convincing explanation.
\end{abstract}



\section{Longmore 4}

Longmore~4 (Lo~4) is a low-surface-brightness planetary nebula (PN), discovered
by Longmore (1977) in the course of the ESO-SRC southern-hemisphere sky survey.
The first spectroscopic observations of its $V=16.6$ central star by M\'endez
et~al.\ (1985) led to a PG~1159$-$035 spectral classification.

The PG~1159-type hydrogen-deficient planetary-nebula nuclei (PNNi) and white
dwarfs have spectra with conspicuous C~$\scriptstyle\rm IV$ and
He~$\scriptstyle\rm II$ absorption, and no Balmer absorption. O~$\scriptstyle\rm
VI$ 3811-3834~\AA\ is often seen in emission. In various spectral-classification
schemes, stars of this type have also  been assigned to types
``O~$\scriptstyle\rm VI$'' or ``O(C).'' Werner et al.\ (2007) have recently
revised the atmospheric parameters for Lo~4 to $T_{\rm eff}=170,000$~K, $\log
g=6$.

The spectrum of Lo~4 is quite similar to that of K~1-16, the first known
pulsating PNN (Grauer \& Bond 1984) and a member of the GW~Vir class of
non-radial pulsators. The close similarity was borne out by the discovery of
pulsations in Lo~4 by Bond \& Meakes (1990). Lo~4 was, at the time, only the
second known pulsating PNN, having a strong periodicity near 31~min
(1850~s), along with at least 8 other pulsation modes ranging from 831 to
2325~s.

\section{A Spectacular Mass-Loss Event in Lo 4}

A transient mass-loss event that occurred in Lo~4 in 1992 was discovered by
Werner et~al.\ (1992, 1993). For several days the spectrum changed from PG~1159
to a [WCE] type ([WC2-3]), having strong emission at the C~$\scriptstyle\rm IV$
complex near 4659~\AA, He~$\scriptstyle\rm II$ 4686~\AA, and O~$\scriptstyle\rm
VI$ 5291~\AA, but then the star reverted to its normal PG~1159 spectrum. The
mass-loss rate during this outburst was estimated at $\log (dM/dt)\simeq-7.3$
(in units of $M_\odot$/yr). As emphasized by Werner et al., this event was a
unique phenomenon, never before seen in any hot post-AGB star.

\section{Spectroscopic Monitoring of Lo 4}

At the time of the 1992 mass-loss event, only a few spectra had ever been taken
of Lo~4. This suggested that the outbursts may not be rare. 

In order to explore
this possibility, I began a program of regular spectroscopic monitoring of Lo~4
in 2003. I am using the 1.5-m telescope of the SMARTS Consortium
(www.astro.yale.edu/smarts), located at Cerro Tololo. The spectrograph setup
gives a wavelength coverage of 3650--5400~\AA, with a resolution of 4.3~\AA\ and
a S/N per resolution element of $\sim$30--40 on good nights.  

\begin{figure}[h]
\includegraphics[width=5.1in]{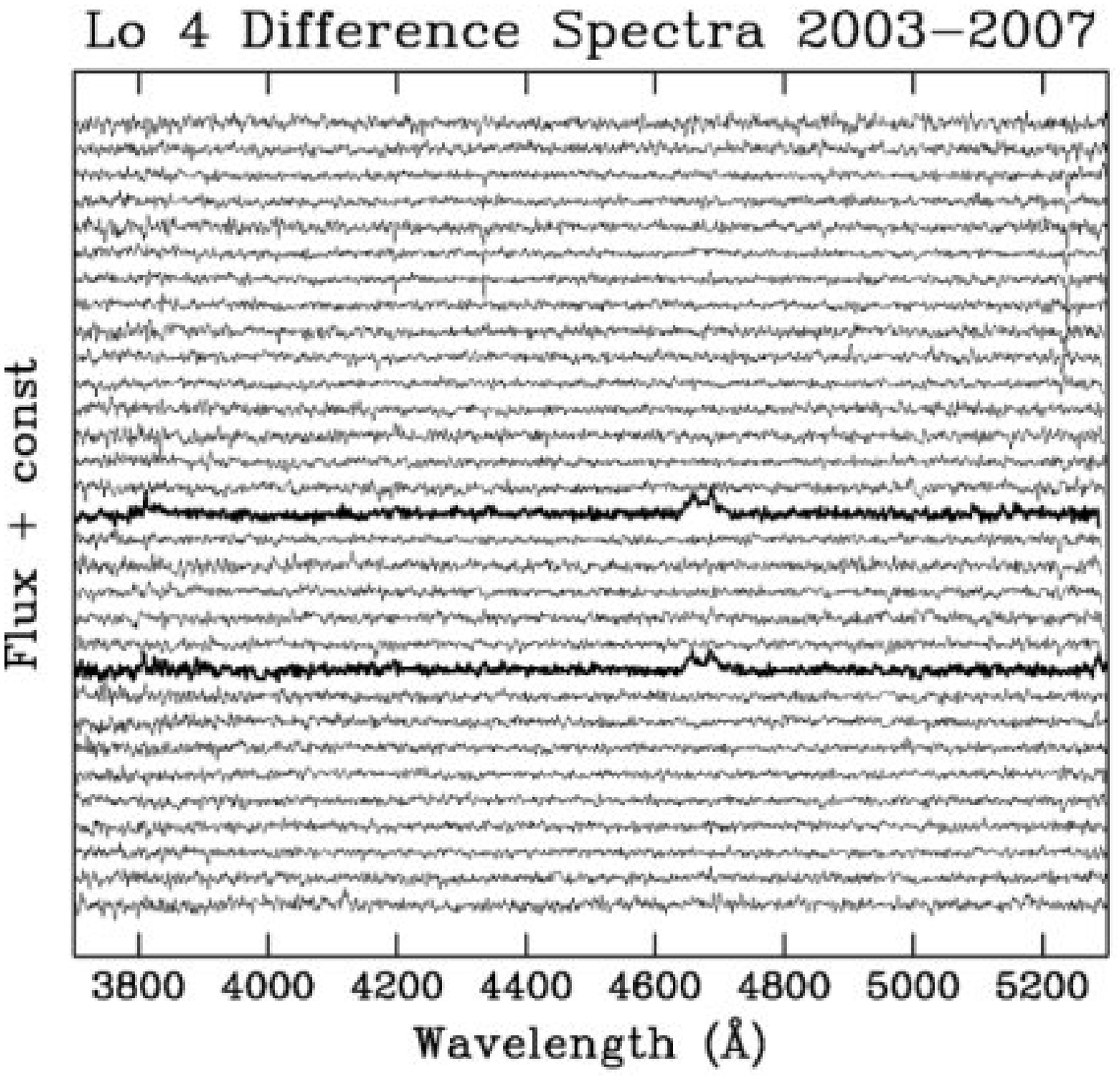}
\begin{caption}
{31 spectra of Lo~4 between 2003 Feb 20 (top) and 2007 Jun 28 (bottom),
obtained with the SMARTS 1.5-m telescope. The mean of all normal  spectra has
been subtracted from each individual spectrum. The normal PG~1159 spectrum was
present on 29 nights. On the nights of 2006 Jan 16 and Nov 30 (plotted with
thick lines) the C~$\scriptstyle\rm IV$+He~$\scriptstyle\rm II$ complex was seen
in emission. Sharp ``absorption'' features are CCD artifacts. Spacing between
spectra is 0.375 of the continuum level.}
\end{caption}
\end{figure}

Between 2003 February and 2007 June, 31 usable spectra were obtained at
essentially random times. Two new mass-loss events were detected during this
interval. Fig.~1 shows the 31 spectra, arranged in time sequence from top to
bottom. In order to highlight variations in the spectra, the mean of all spectra
has been subtracted from each individual spectrum. The resulting spectral
differences show only residual noise, except on the two nights whose spectra are
plotted with thick lines. On these two nights (2006 Jan 16 and Nov 30) Lo~4 was
in the [WCE] state first discovered by Werner et al.\ in 1992.

Fig.~2 shows the spectra in more detail. The top spectrum is the mean of all 29
observations in the PG~1159 state, showing the usual absorption lines of
C~$\scriptstyle\rm IV$ and He~$\scriptstyle\rm II$, and the emission doublet of 
O~$\scriptstyle\rm VI$ 3811-3834~\AA\null. During the two outburst nights in
2006 Jan and Nov, however, C~$\scriptstyle\rm IV$ and He~$\scriptstyle\rm II$
have gone into strong emission. The O~$\scriptstyle\rm VI$ 3811-3834~\AA\ lines
also appear to be stronger, and the 2006 Nov 30 spectrum shows that
O~$\scriptstyle\rm VI$ 5291~\AA\ has gone into emission as well.  Both of these
spectra are thus quite similar to that seen in 1992 by Werner et al.  As noted
in Fig.~2, the previous observations, taken 53--56 days earlier, had shown the
normal PG~1159 spectrum. The next observations, taken 15 and 17 days later,
showed that the star had reverted to the PG~1159 state.

\begin{figure}[h]
\includegraphics[width=5.25in]{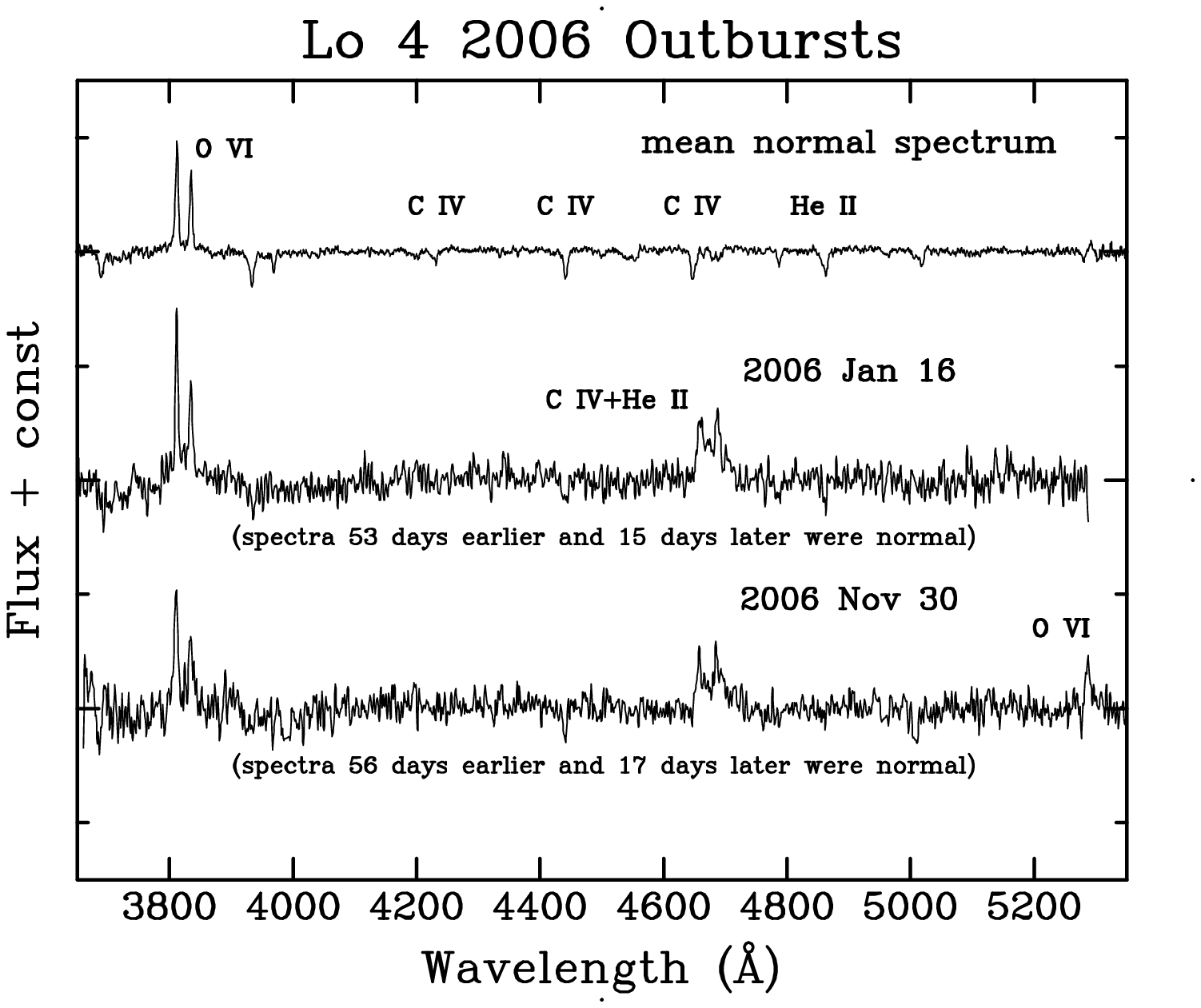}
\begin{caption}
{The mean of 29 normal spectra of Lo~4 is shown at top. On 2006 Jan 16 (middle)
and Nov 30 (bottom) the star was in its transient [WCE] stage, with strong
C~$\scriptstyle\rm IV$ and He~$\scriptstyle\rm II$ emission. Spectra taken
53--56 days earlier, and 15--17 days later, showed the normal PG~1159 spectrum.
Tick marks on the $y$ axis are spaced at 0.5 of the continuum level.
}
\end{caption}
\end{figure}

\section{Constraints and Speculations}

At present there seems to be no entirely satisfactory explanation for these
spectacular, transient mass-loss episodes.  One constraint is that the transient
stellar wind is {\it hydrogen-deficient}, showing that the outbursts are
unlikely to be related to an accretion event involving a companion star, debris
disk, infalling planet, or other exotic external cause. The short durations of
the episodes indicate that they are not driven on an evolutionary timescale.

I have also been monitoring the spectroscopically very similar, pulsating PNN
NGC~246. To date, it has not shown any events like those of Lo~4.

Lo~4 does lie in the HR diagram for hot post-AGB remnants near the boundary
between [WCE] and PG~1159 objects. This suggests that it may take only a small
perturbation to move the star back into the [WCE] stage. I thus speculate that
occasionally many pulsation modes may be in phase, giving a temporarily large
surface-temperature amplitude, which might trigger the outbursts. If so, we may
expect the outbursts to occur periodically, at the beat period between the
pulsation modes.

Three mass-loss episodes have now been seen in the $\sim$40 spectra of Lo~4 that
have ever been obtained. Thus the star is in its ``high'' state $\sim$8\% of the
time. The time-averaged mass-loss rate from the outbursts is 
$\sim$$0.08\times10^{-7.3}\simeq4\times10^{-9}\,M_\odot$/yr. Since the mass-loss
rate during Lo 4's normal PG~1159 state is
$\sim$3--$30\times10^{-9}\,M_\odot$/yr (Werner, priv.\ comm., based on {\it FUSE}
observations), the outbursts will have little additional effect on the evolution
of the star.

These abrupt transitions between the [WCE] and PG~1159 states support the
commonly accepted evolutionary link between the two types. Further observations
could test whether the outbursts occur periodically (at the beat period), and
other PNNi lying near the [WCE]--PG~1159 transition should be monitored.
 
\acknowledgements 
I thank Fred Walter for telescope scheduling and the excellent SMARTS
service observers for obtaining the spectra.



\begin{thebibliography}{}

\bibitem[]{} Bond, H. E., \& Meakes, M. G. 1990, AJ, 100, 788
\bibitem[]{} Grauer, A. D., \& Bond, H. E. 1984, ApJ, 277, 211
\bibitem[]{} Longmore, A. J. 1977, MNRAS, 178, L251
\bibitem[]{} M\'endez, R. H., Kudritzki, R. P., \& Simon, K. P. 1985, A\&A, 142,
289
\bibitem[]{} Werner, K., et al. 1992, A\&A, 259, L69 
\bibitem[]{} Werner, K., et al. 1993, in Planetary Nebulae, IAU Symp. 155, 494
\bibitem[]{} Werner, K., Rauch, T., \& Kruk, J.~W. 2007, A\&A, 474, 591 

\end{thebibliography}
\end{document}